\documentclass[pra,showpacs,twocolumn,aps]{revtex4-1}

\usepackage[utf8]{inputenc}
\usepackage{latexsym}
\usepackage{amssymb}
\usepackage{amsfonts}
\usepackage{epsfig}
\usepackage{psfrag}
\usepackage{graphicx}

\usepackage{epsfig,color}
\usepackage{psfrag}
\usepackage[utf8]{inputenc} 
\definecolor{darkgreen}{rgb}{0,0.4,0}

\newcommand{\bea}{\begin{eqnarray}}
\newcommand{\ea}{\end{eqnarray}}
\newcommand{\eea}{\end{eqnarray}}

\usepackage{amsmath}
\usepackage{longtable}
\usepackage{color}

\begin{document}

\title{When Repulsion Creates Pairing: A New Perspective on Unconventional Superconductivity}
\author{Patrick Navez}
\affiliation{Laboratoire Charles Coulomb UMR 5221 CNRS-Universit\'e de Montpellier, F-34095 Montpellier, France
}

\date{\today}
\begin{abstract}
Despite the repulsive Coulomb law and the Pauli statistics that do not favor bound states, attraction between electrons or holes is nevertheless possible in the context of many body interaction and of valley potential landscapes, reminiscent of exotic superconducting materials. In particular, in 1965, Kohn and Luttinger published a note revealing that the dynamical screening of the repulsive Coulomb interaction leads, under certain conditions, to an effective attraction necessary for the formation of Cooper pairs. We propose such a formalism adapted to the cuprates, where the screening arises from the superexchange dynamics of virtual holes in the oxygen orbitals of the $Cu O_2$ plane. Inspired from the Bardeen-Copper-Schrieffer (BCS) theory, we can derive some predictions for the temperature-doping phase diagram (pseudo-gap, strange metal, antiferromagnetism, superconducting, and normal states)  in semi-quantitative agreement with observations.
\end{abstract}
\maketitle

\section{Introduction}

Discovered in the beginning of the twentieth century, superconductivity  remains one of the most fascinating phenomenon in physics. It offers a unique opportunity to observe, control, and exploit the indefinite flow  of the charge carriers without energy loss. Despite decades of experimental investigation, no measurable current damping has been observed inside a low temperature superconducting material. This remarkable behavior is extremely specific to any general superfluid, also encountered in many body systems such as helium or ultra-cold atoms
\cite{NAVEZ2005241,lirias1912221}.


The fundamental question is what allows a fluid to flow indefinitely without friction, even in the presence of obstacles such as container walls or impurities. To date, no complete explanation of this phenomenon has been established. One important clue, however, lies in the possibility of Bose–Einstein condensation of the particles that carry the flow. If these particles are bosons, then below a critical temperature a fraction of them condenses into a single macroscopic quantum state, sharing the same quantum wave function that governs the superfluid dynamics.


In a superconductor, electrons can enter such a collective quantum state only if they first pair up. Each pair then behaves like a boson, allowing many pairs to occupy the same quantum state. At first sight, this seems surprising because electrons naturally repel each other through the Coulomb interaction. The key lies in the crystal lattice of the material. As electrons move through the lattice, they slightly displace the surrounding positively charged ions, creating an effective attractive interaction between electrons. This attraction can overcome their natural repulsion and bind them into so-called Cooper pairs. The existence of these pairs has been experimentally confirmed in most conventional superconducting metals, such as aluminum, mercury, and niobium, providing strong support for the Bardeen–Cooper–Schrieffer (BCS) theory, whose authors were awarded the Nobel Prize in Physics in 1972.


\begin{figure}
 \centering
 \includegraphics[width=8cm]{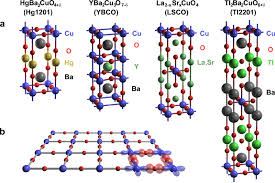}
 \caption{a) Family of cuprates containing the $CuO_2$ plane shown in b). The orbitals $p$ for oxygen are represented in red and $d$ for copper in blue. Pictures taken from \cite{doi:10.1073/pnas.1301989110}.}
 \label{fig:0}
\end{figure}

The BCS framework remained the cornerstone of superconductivity research until the discovery, in 1986, of a new class of ceramic materials exhibiting superconductivity at temperatures far higher than originally expected \cite{osti_1357580}, well above the boiling point of liquid nitrogen. The first of these materials to be extensively studied were the cuprates, which contain layers of $CuO_2$ planes separated by various electropositive atoms such as Hg, Ba, Sr, and Y (see Fig.\ref{fig:0}). Since then, other families
have joined the group of unconventional
superconductors, including nickelates, iron-based superconductors, and twisted bilayer graphene.

Unlike ordinary metals, these materials are often poor conductors or even insulators. Their ability to become superconducting therefore presents a major challenge to our understanding of condensed-matter physics. Despite four decades of intensive research, the microscopic origin of pairing in these materials remains unresolved. Experimental observations, including the weak isotope effect in many cuprates (i.e. influence of different isotopes of atoms such as mercury on attraction), suggest that lattice vibrations play only a limited role in generating the attractive interaction required for superconductivity.


\begin{figure}[!h]
\vspace*{0.5cm}
\hspace*{-8cm}{\bf a)}
\begin{picture}(1000,50)(0,0)%
\includegraphics[width=8cm]{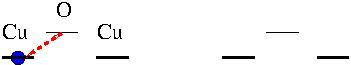}%
\put(-200,-20){ $\hspace{-0.5cm} E_{1}= -\displaystyle \frac{t_{pd}^2}{\epsilon_p }  \hspace{4cm} E_0= 0$}
\put(-230,-80){
\includegraphics[width=8cm]{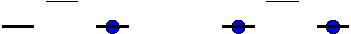}}%
\put(-200,-100){ $\hspace{-0.5cm}  E_{1'}=-\displaystyle \frac{t_{pd}^2}{\epsilon_p } \hspace{3cm} E_2=  -
\displaystyle \frac{ 2 t_{pd}^2}{\epsilon_p +U_{pd}}$}
\end{picture}%
\vspace*{4cm}
\hspace*{-8cm}{\bf b)}
\centering
\begin{picture}(1000,50)(0,0)%
\includegraphics[width=8cm]{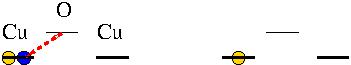}%
\put(-200,-20){ $\hspace{-1cm} E_{1}=U_d+\displaystyle  \frac{2t_{pd}^2}{U_d-\epsilon_p-U_{pd}} \quad \quad \quad \quad\quad \quad E_0= -\frac{t_{pd}^2}{\epsilon_p }$}
\put(-230,-80){
\includegraphics[width=8cm]{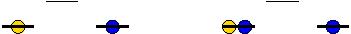}}%
\put(-200,-100){ $\hspace{-1cm}  E_{1'}=-\displaystyle \frac{2t_{pd}^2}{\epsilon_p +U_{pd}} \quad \quad   E_2= U_d -
\displaystyle \frac{ t_{pd}^2}{\epsilon_p +2 U_{pd}}  + \frac{2 t_{pd}^2}{U_d-\epsilon_p - 2U_{pd}}$}
\end{picture}%
\vspace*{4cm}
\caption{{\bf Schematic representation of pairing:}
\\
In a) and b), four independant triads consisting each of two $d$ sites interacting via a $p$ site containing holes in blue and gold colors. Comparison between the energy of two single holes  $E_1 +E_{1'}$ and the energy of no hole and pairing of energy $E_0+E_2$.
Energy graph of $E_\nu$ for various occupation $\nu$ for two cases:
\\
{\bf a) }
In absence of additional hole, pairing is not possible, since
$E_1 + E_{1'} -(E_0 + E_{2})= \displaystyle \frac{2t_{pd}^2}{\epsilon_p }-\frac{2t_{pd}^2}{\epsilon_p +U_{pd}} <0$
\\
{\bf b) } In presence of one assisting hole in gold color, pairing becomes possible for high $U_d$:
\\
$E_1 + E_{1'} -(E_0 + E_{2}) \stackrel{U_d \rightarrow \infty}{=} \displaystyle \frac{t_{pd}^2}{\epsilon_p }-\frac{2t_{pd}^2}{\epsilon_p +U_{pd}}+\frac{ t_{pd}^2}{\epsilon_p +2 U_{pd}}  > 0$
\\
For very high on-site repulsion $U_d$, the repulsive off-site potential $U_{pd}$ is crucial for pairing.
Its screening role reduces the  delocalisation of any hole on the $p$ orbital and increases the energy. Such a ``thin'' hole contrasts with a ``fat'' hole with a more delocalized state on the $p$ orbital.
}
\label{fig:2}
\end{figure}

As a result, the electrons—or, equivalently in many descriptions, the holes—must somehow develop an effective attraction despite their underlying Coulomb repulsion. In a seminal paper published in 1965, Kohn and Luttinger \cite{KL} proposed that the dynamical screening of the Coulomb interaction by surrounding electrons could under certain conditions generate an effective attraction between particles at  certain ranges \cite{LIU199781,Belyavsky}.
In a simple picture, the other surrounding electrons obeying the Pauli exclusion principle contribute to alter the profile of the two-body interaction.
In cuprates, charge carriers move primarily within the two-dimensional $Cu O_2$ planes, navigating a  landscape of valleys and mountains formed by copper and oxygen orbitals. The same quantum processes responsible for antiferromagnetic superexchange interactions \cite{Anderson,ZR,ZR2} also govern the motion of these carriers. This naturally raises an intriguing question: can the screening mechanism identified by Kohn and Luttinger combine with superexchange dynamics to generate an effective attraction between holes within these planes?
Recent work suggests that the answer is yes \cite{bcsnavez,doi:10.1126/sciadv.abh2233,doi:10.1073/pnas.2117735119}. In this article, we present an intuitive description of this mechanism, showing how a fundamentally repulsive interaction can give rise to attraction under appropriate conditions. The discussion relies only on concepts from quantum mechanics typically encountered at the graduate level.

\section{The triad model}

The many-body dynamics of charge carriers in the $CuO_2$ plane is extremely difficult to describe exactly, even with modern computational resources. One of the standard theoretical frameworks is the Fermi–Hubbard model \cite{PhysRevB.44.7504,PhysRevB.71.134527,PhysRevB.53.8751,PhysRevB.45.7959,PhysRevLett.128.087001}, in which electrons (or holes) move between neighbouring copper sites while experiencing a strong Coulomb repulsion whenever two particles occupy the same site. Despite decades of effort, no exact solution is known for this model, and even sophisticated numerical calculations often provide limited physical insight into the origin of superconducting pairing \cite{PhysRevResearch.2.023172,Weber_2010,Baeriswyl_2009,doi:10.1126/science.235.4793.1196,PhysRevLett.60.2430,PhysRevB.81.064515}.
Within the $CuO_2$ plane, oxygen atoms are typically found in the $O^{2-}$ ionic state, with their $2p$ (or $p$) orbitals essentially filled by pairs of electrons with opposite spins. Copper atoms, by contrast, are close to the $Cu^{2+}$ state, with the orbital $3 d_{x^2-y^2}$ (or $d$) lying near half-filling (see \ref{fig:0}b).
To understand how pairing might arise, it is sufficient to focus on a minimal system consisting of two neighboring copper atoms interacting through a single oxygen atom. This three-site "triad" model captures the essential physics while remaining mathematically accessible, requiring only elementary second-order perturbation theory.



Because the oxygen orbitals are already filled, it is more convenient to describe the system in terms of holes rather than electrons. A hole can be thought of simply as the absence of an electron. Each orbital can host at most two holes with opposite spins, in accordance with the Pauli exclusion principle.
Let $\epsilon_p$ denote the energy difference between oxygen and copper orbitals, and $U_d$ and $U_p$ the on-site Coulomb repulsion energies for each orbital, $U_{pd}$ the inter-orbital Coulomb interaction, and $t_{pd}$ the hopping amplitude between neighboring orbitals.

Using second-order perturbation theory in the small parameter $t_{pd}$, we calculate the ground-state energies for configurations containing no hole $E_0$, one hole $E_1$ and $E_{1'}$, and two holes at different sites $E_2$. The general expression for the energy with its second order correction is given by:
\begin{eqnarray}
E_i \simeq E_i^{(0)} + \sum_{i'}\frac{t_{pd}^2}{E_{i}^{(0)}- E_{i'}^{(0)}}
\end{eqnarray}
where the summation refers to all possible virtual transition states of energy $E_{i'}^{(0)}$ labelled by $i'$.
The physical interpretation is straightforward. A hole virtually hops from a copper orbital $d$  to a neighboring oxygen orbital $p$ and then return. The lower the energy cost of the virtual state, the stronger the hybridization or delocalization effect. As illustrated in Fig.\ref{fig:2}a), an isolated hole lowers the total energy as much as the energy transition $E_{i'}^{(0)}=\epsilon_p$ is smaller.  For virtual transitions involving two single occupancies however, the shift in energy $\epsilon_p+ U_{pd}$ is higher due to the Coulomb repulsion but globally less than the case of an isolated hole.

Pairing occurs if the presence of two holes is more favorable energetically than two isolated holes relative to the background energy as shown in Fig.\ref{fig:2}. The resulting attraction energy should be negative:
\begin{eqnarray}\label{att}
U= E_2 +E_0-E_1-E_{1'} < 0\, .
\end{eqnarray}
This energy is positive for \ref{fig:2}a) and excludes pairing.
The situation changes when neighbouring holes are present nearby.
They  generate a screening that modifies the virtual hopping processes in such a way that an effective attraction can emerge. In this case, calculations are more involved but easily tractable.
For example in the last triad containing  three particles of Fig.\ref{fig:2}b), $E_{2}^{(0)}=U_d$ and three transition energies $E_{i'}^{(0)}=\epsilon_p +2U_{pd}$ and $E_{i'}^{(0)}=\epsilon_p +2U_{pd} +U_d$ counted twice contribute to the second order correction. As a result, the transition energy is reduced by the presence of repulsion energy $U_{pd}$ leading to a global net attraction.
In the limit of strong on-site repulsion $U_d$, the effect becomes particularly pronounced with a quadratic contribution in  $U_{pd}$.
Only in the case of two added holes that is left as an exercise for students, attraction occurs with a linear contribution in $U_{pd}$ as shown in \cite{bcsnavez}.

We conclude that, although the underlying Coulomb interaction remains repulsive, the many-body environment reshapes the energy landscape making pairing energetically favorable.
A useful picture is to compare "thin" and "fat" holes. A thin hole remains localized mainly on a copper orbital, whereas a fat hole spreads more strongly over neighbouring oxygen orbitals.
The greater this localization, the stronger the resulting attraction can become. In this sense, the surrounding charges assist the pairing process by modifying the spatial extent of the hole wave functions.

\begin{figure}[!h]
\centering
\vspace*{1cm}
\hspace*{-8cm}
{\bf a)}
\vspace*{0.5cm}
\begin{picture}(1000,50)(0,0)
 \put(20,-100){\includegraphics[width=7cm]{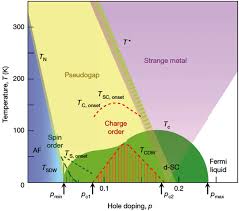}}
 \end{picture}
 \vspace*{0cm}
 \begin{picture}(1000,50)(0,0)
 \put(0,-40){\bf  b)}
 \put(0,-255){\includegraphics[width=9cm]{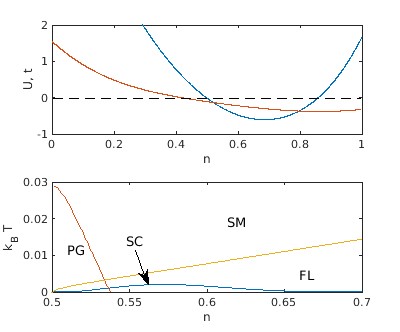}}
 \end{picture}
 \vspace*{8cm}
\caption{{\bf Comparison between experiment and theory : a)} Typical experimental phase diagram of temperature versus hole doping $p$ for the copper oxides, indicating where various phases. Graph taken from  \cite{osti_1357580};
\\
{\bf b)}
 First theory graph of the hopping factor $t$ between effective $d$ orbitals in red and attraction potential $U$ in blue vs. the density $n=(p+1)/2$ per spin (or doping $p$). The dashed line on the x axis is a guide for the eye.  Second theory graph of the critical superconducting (SC) temperature vs. the doping $n$  and of the transition lines for the pseudogap (PG) the strange metal (SM) and the Fermi liquid (FL) phases.
 Parameter values: $\epsilon_p=0.4$, $U_{pd}=1.55$, $U_d=6.6$ (in eV unit) and with an additional hopping parameter between the $p$ orbitals $t^{eff}_{pp}=0.038$ explained in \cite{bcsnavez}.}
 \label{fig:3}
\end{figure}

\section{Extension to the $CuO_2$ lattice}

The simple reasoning obtained so far from three sites
can be generalized to the full $CuO_2$ plane lattice. Inspired from the general BCS approach where a virtual phonon process results in an effective  attraction, a virtual process involving the sublattice of the oxygen $p$ orbital results in an effective attraction of the holes between neighbouring copper $d$ orbitals.

In \cite{bcsnavez}, we have merged both well-established concepts of screening developed by Kohn and Luttinger \cite{KL} and the superexchange developed by Anderson \cite{Anderson} into one unified theory which is a mere elegant use of the Schrieffer-Wolff technique applied to the three-band Hamiltonian model to cuprates.
We thus obtain a mean-field model in a basis of states that incorporates correlations between the $p$ and $d$ orbitals. The attractive interaction in Eq. (\ref{att}) corresponds to the weighted average of the interaction energies associated with the three screening configurations of the triad model, namely those involving zero, one, or two holes.


The half-filling regime of the hybridized $d$ orbitals is not in the  superconducting phase  but in the antiferromagnetic phase (AF) shown in Fig.\ref{fig:3}. It corresponds to exactly one hole per site
whose spin direction alternates between two $d$ sites. By substituting in the cuprates some atoms with others of lower valences  (e.g. $La$ of valence $3+$ with portion $x$ of $Sr$ of valence $2+$ in  $La_{2-x}Sr_x Cu O_4$), the  sites contain extra doping holes of exact concentration  $p=x$ that triggers an attraction to the superconducting phase (SC). Note that the effective hopping $t$ between sites depends also on the doping.
Our approach provides also some insights on transitions to the enigmatic pseudogap (PG) and strange metal (SM) phases to be distinguished from the normal metal Fermi liquid (FL) phase \cite{bcsnavez}. These exotic phases will not be explained here but illustrate the complex properties of such materials.

The resulting formalism  successfully provides many explanations for the behavior of the cuprates and therefore appears robust for other applications in solid-state theory. It suggests also that room temperature superconductivity is possible  for significant screening in superexchange interactions and a lower energy gap $\epsilon_p$ \cite{doi:10.1073/pnas.2207449119}.
Additional experimental evidences - such as transport properties, specific heat measurements, charge or spin order \cite{doi:10.1073/pnas.1910411116} or nematicity \cite{doi:10.1073/pnas.2206481119} - are still needed to fully validate the relevance of the present approach.

\section{Summary}

Can electrons or holes ``fall in love'' with one another?

\begin{figure}
 \centering
 \includegraphics[width=8cm]{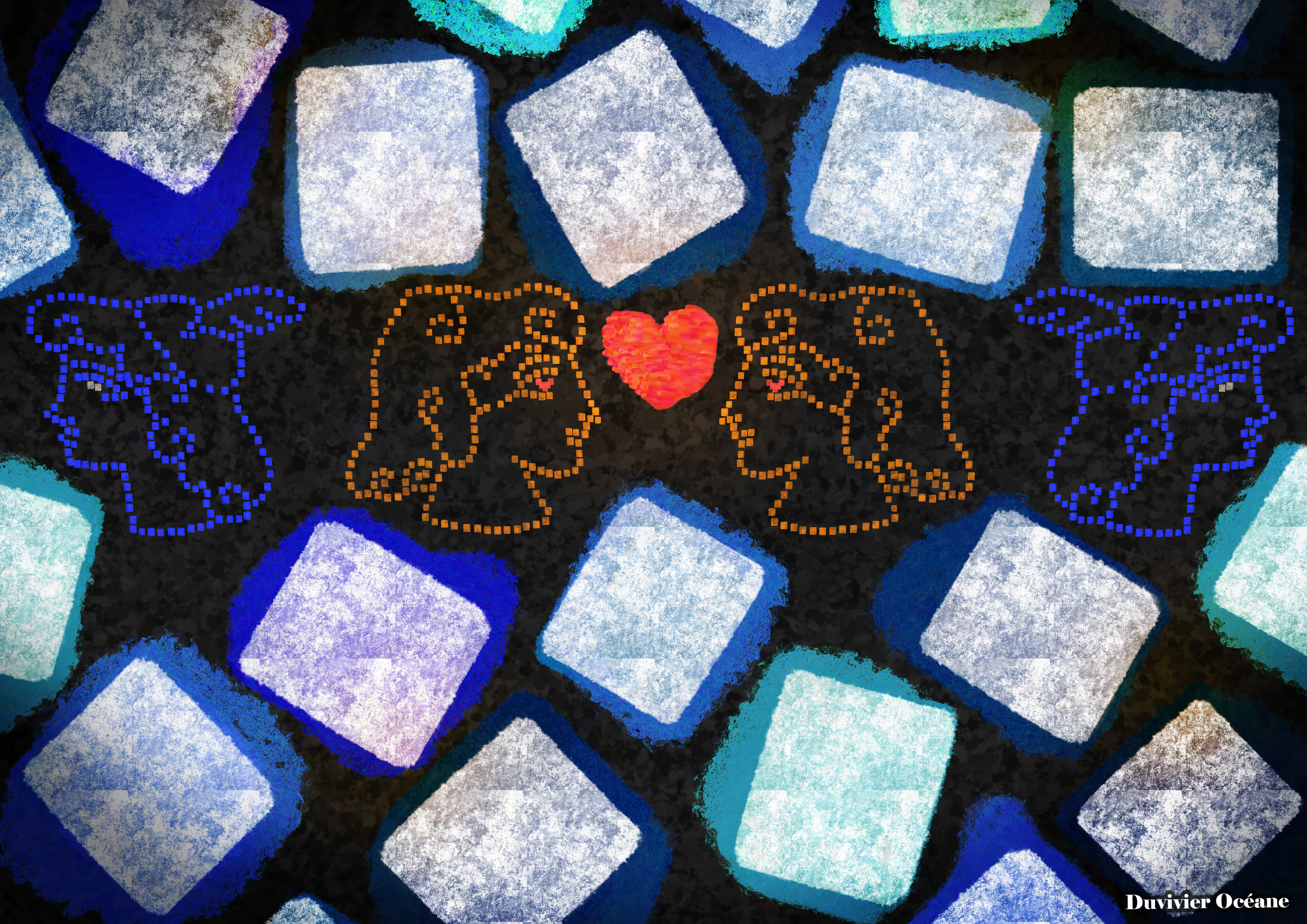}
 \caption{Attractive repulsion represented in Minoan style.
  The mosaics symbolize electrons surrounding holes with human face. The red holes are depicted with human faces turned toward each other, illustrating their effective attraction induced by the repulsion from the surrounding blue holes.}
 \label{fig:4}
\end{figure}

Surprisingly, the answer is ``yes'' provided that other charges are present to localize the wave function in order to transform a repulsion into an attraction  (see Fig.\ref{fig:4}). An everyday analogy would be someone in love with a woman and who is asking his friend(s) to be unpleasant with her and thus so repulsive that, as a result, she can inadvertently fall into his arms and become attracted to him.


This unexpected mechanism illustrates how familiar physical concepts such as Coulomb repulsion, screening, and superexchange can combine to produce entirely new phenomena.
Whether this mechanism fully explains superconductivity in the cuprates remains an open question. Nevertheless, it offers a promising framework that deserves further theoretical development and experimental investigation, both in cuprates and in other families of unconventional superconductors.


{\bf Acknowledgements:}
PN
is grateful to André-Marie Tremblay, Joseph Betouras, Xenophon Zotos, Ioannis Rousochatzakis, and Todor Mishonov
for fruitful discussions. A special thanks to Océane Duvivier for the illustration in Fig.\ref{fig:4}.

\bibliographystyle{apsrev4-1}
\bibliography{paperrefbcs}

\end{document}